\documentclass[10pt,conference]{IEEEtran}
\IEEEoverridecommandlockouts
\usepackage{cite}
\usepackage{amsmath,amssymb,amsfonts}
\usepackage{graphicx}
\usepackage{textcomp}
\usepackage{xcolor}
\usepackage[export]{adjustbox}
\usepackage{array, makecell, cellspace}

\usepackage{threeparttable}

\usepackage[hyphens]{url}

\usepackage{multirow}
\usepackage[ruled]{algorithm2e}
\usepackage{subfigure}		
\usepackage{tikz}

\usepackage{xcolor}

\usepackage{xcolor,pifont}
\newcommand*\colourcheck[1]{%
  \expandafter\newcommand\csname #1check\endcsname{\textcolor{#1}{\ding{52}}}%
}
\colourcheck{blue}
\colourcheck{green}
\colourcheck{red}

\newcommand*\colourx[1]{%
  \expandafter\newcommand\csname #1x\endcsname{\textcolor{#1}{\ding{54}}}%
}
\colourx{red}

\def\BibTeX{{\rm B\kern-.05em{\sc i\kern-.025em b}\kern-.08em
    T\kern-.1667em\lower.7ex\hbox{E}\kern-.125emX}}

\usepackage{caption}
\usepackage[margin=0.6in]{geometry}

\renewcommand{\IEEEbibitemsep}{0.71pt plus 1pt}
\makeatletter
\IEEEtriggercmd{\reset@font\normalfont\fontsize{7.5pt}{4.2pt}\selectfont}
\makeatother
\IEEEtriggeratref{1}

\begin{document}

\title{SeqL: Secure Scan-Locking for IP Protection\\
}

\author{\IEEEauthorblockN{Seetal Potluri}
\IEEEauthorblockA{\textit{Department of ECE} \\
\textit{North Carolina State University}\\
Raleigh, U.S. \\
spotlur2@ncsu.edu}
\and
\IEEEauthorblockN{Aydin Aysu}
\IEEEauthorblockA{\textit{Department of ECE} \\
\textit{North Carolina State University}\\
Raleigh, U.S. \\
aaysu@ncsu.edu}
\and
\IEEEauthorblockN{Akash Kumar}
\IEEEauthorblockA{\textit{Department of Computer Science} \\
\textit{Technical University of Dresden}\\
Dresden, Germany \\
akash.kumar@tu-dresden.de}
}

\maketitle

\begin{abstract}
Existing logic-locking attacks are known to successfully decrypt {\em functionally correct key} of a locked combinational circuit. It is possible to extend these attacks to real-world Silicon-based Intellectual Properties (IPs, which are sequential circuits) through scan-chains by selectively initializing the combinational logic and analyzing the responses. 
In this paper, we propose {\em SeqL}, which achieves functional isolation and locks selective flip-flop functional-input/scan-output pairs, thus rendering the decrypted key {\em functionally incorrect}. 
We conduct a formal study of the scan-locking problem and demonstrate automating our proposed defense on any given IP. We show that {\em SeqL} hides functionally correct keys from the attacker, thereby increasing the likelihood of the decrypted key being {\em functionally incorrect}. When tested on pipelined combinational benchmarks (\texttt{ISCAS}, \texttt{MCNC}), sequential benchmarks (\texttt{ITC}) and a fully-fledged \texttt{RISC-V CPU}, {\em SeqL} gave 100\% resilience to a broad range of state-of-the-art attacks including SAT~\cite{pramod:host15}, Double-DIP~\cite{double-dip}, HackTest~\cite{hacktest},  SMT~\cite{smt-attack},  FALL~\cite{pramod-faa},  Shift-and-Leak~\cite{shift-and-leak-iccad19} and Multi-cycle attacks~\cite{guin:2018}. 
\end{abstract}

\begin{IEEEkeywords}
IP Piracy, Logic Locking, Scan-chains
\end{IEEEkeywords}

\vspace{-.25em}
\section{Introduction}
\vspace{-0.25em}
\label{sec:introduction}
\noindent

Logic-locking is a solution that was touted to address IP piracy threats in the semiconductor supply chain. 
This technique adds key-gates with one input driven by secret key, to obfuscate IP's inner details. The transformation is reversed only upon application of the programmed secret key, thus preserving the IP's original function. Unfortunately, logic-locking has been a cat-and-mouse game where existing locking proposals~\cite{farinaz:epic, jv:tc15,dupuis:iolts14,TTLock,SARLock,Xie:ches2016,SFLL} fail to ever-advancing attacks~\cite{pramod:host15,double-dip,hacktest,smt-attack,pramod-faa}. 
Although these attacks primarily target combinational circuits, they can be extended to real-world sequential circuits through \emph{scan-chains}.
But the fundamental attack assumption is that inputs are controllable and outputs are observable. Thus, if the scan-chains are secured, it would be possible to provide a secure logic locking solution. 

This paper proposes \emph{SeqL}, a new logic locking technique that secures scan-chains. SeqL advances the prior work on design-for-security (DFS)~\cite{fortis, guin:2018,rajit:encryptFF}, by conducting a formal study and empirically validating the security against a broad class of state-of-the-art attacks.
Although attacks on large-scale sequential designs through functional execution is an open problem, attacks through the scan-chains currently exist. Thus, SeqL serves as the proper first line of defense. We demonstrate how to automate SeqL and quantify its low overheads on large-scale circuits. Therefore, SeqL addresses both the security and practicality challenges of logic locking.

\begin{figure}[!t]
\vspace{-0.5em}
\centering
\includegraphics[scale=0.55]{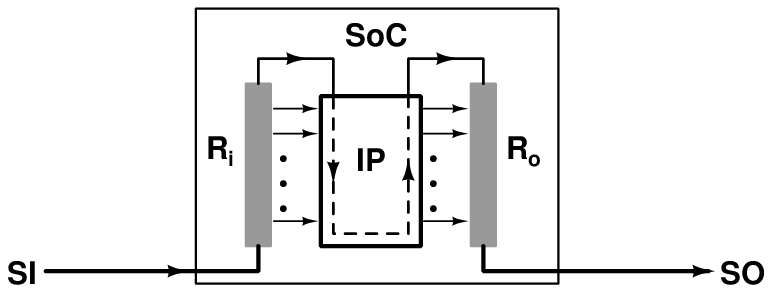}
\vspace{-.55em}
\caption{Scan-based IP access for logic locking attacks. 
}
\label{fig:ip-access}
\vspace{-1.35em}
\end{figure}

Figure~\ref{fig:ip-access} outlines the system we consider. We highlight a simple setting for ease of explanation and elaborate later in our threat model. 
We assume that the primary inputs and primary outputs of the IP under consideration are not accessible, while only the scan-chains are accessible to the attacker, in embedded deterministic test (EDT)-bypass mode. The input register $R_i$ applies primary inputs to the IP, and the output register $R_o$ stores the primary outputs. The scan-chain connects all flip-flops in $R_i$, subsequently to the flip-flops internal to the IP and finally the flip-flops in $R_o$. The scan-input ($SI$) and scan-output ($SO$) ports are controllable and observable respectively by the attacker. Hence, the attacker can apply selective inputs to the IP and observe corresponding IP responses through these ports. 
\begin{figure*}[!t]
\centering
\includegraphics[scale=0.155]{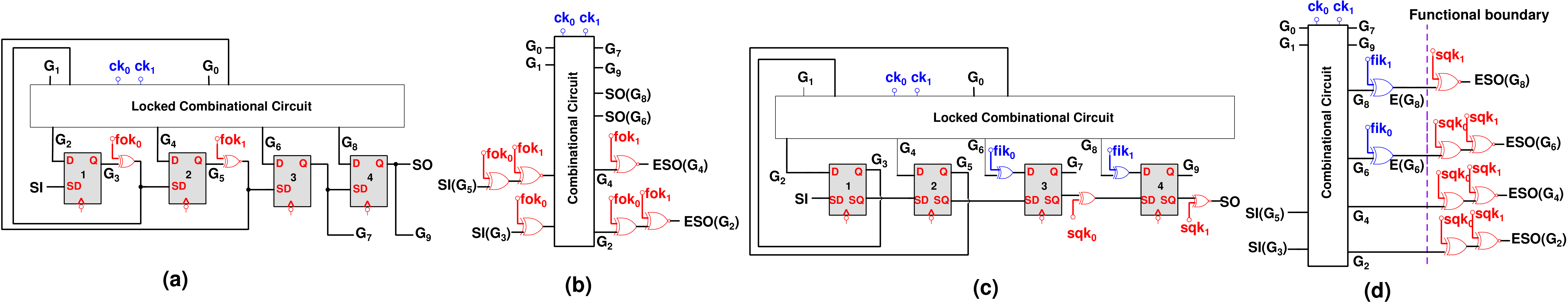}
\caption{(a) {\em EFF}-style: Sample sequential circuit with logic locking and FO locking; (b) Scan-unrolled equivalent of Fig.~\ref{fig:eff-seql-scansat}(a); (c) {\em SeqL}-style: Functional isolation and locked FIs/SQs; and (d) Scan-unrolled equivalent of Figure~\ref{fig:eff-seql-scansat}(c). {\em SeqL} considers flip-flops without feedback.
}
\label{fig:eff-seql-scansat}
\end{figure*}
The main contributions of this paper are as follows:
\begin{enumerate}
	\item We identify there is $100\%$ correlation between flip-flop input (FI) locking and functional output corruption; 
	\item Exploiting this property, we propose {\em SeqL}, that: (a) isolates functional path from the locked scan path; (b) locks FIs and causes functional output corruption; 
	\item {\em SeqL} hides majority of the scan-correct keys which are functionally correct, thus maximizing the probability of the decrypted key being {\em functionally incorrect}; 
	\item The security of {\em SeqL} is also empirically evaluated and verified against a broad set of best known attacks;  
	\item The small overheads of {\em SeqL} and its ease of implementation makes it attractive for industry practice.
\end{enumerate}


\section{Prior Work}
\label{sec:prior}

The first wave of logic locking techniques~\cite{farinaz:epic, dupuis:iolts14, jv:tc15} have been shown to be vulnerable to Boolean Satisfiability (SAT) attack~\cite{pramod:host15}. In SAT-attack, distinguishable input patterns (DIPs) are obtained from the locked circuit and incorrect keys are pruned-off based on oracle's responses to the DIPs. 
Several defenses were then proposed to mitigate SAT-attack, such as {\em Anti-SAT}~\cite{Xie:ches2016}, {\em SARLock}~\cite{SARLock} and {\em Cyclic Obfuscation}~\cite{cyc-obfuscation}, but they have failed to address the vulnerability to {\em AppSAT}~\cite{AppSAT}, {\em Double-DIP}~\cite{double-dip}, {\em CycSAT}~\cite{cycsat}, {\em HackTest}~\cite{hacktest}, {\em BeSAT}~\cite{besat} and machine-learning~\cite{ml-attack} attacks. 
While~\cite{cycsat-resolvable} proposes a new cyclic logic locking technique to defend {\em CycSAT}~\cite{cycsat}, {\em TTLock}~\cite{TTLock} and Secure Function Logic Locking (SFLL)~\cite{SFLL} were the only locking schemes that were broadly resilient to these attacks, yet they recently failed against functional analysis of logic locking (FALL)~\cite{pramod-faa} and SMT~\cite{smt-attack} attacks.

Additionally, to address the issue of defending against SAT-attack on sequential circuits, several DFS techniques have been proposed: (1) {\em FORTIS}~\cite{fortis}, (2) Robust DFS ({\em RDFS})~\cite{guin:2018} and (3) Encrypt Flip-Flop ({\em EFF})~\cite{rajit:encryptFF}. 
{\em FORTIS}~\cite{fortis} is vulnerable to multi-cycle-test attacks~\cite{guin:2018} ; {\em RDFS}~\cite{guin:2018} addresses these issues but necessitates routing of a global {\em Test} signal to all the key-based scan flip-flops, adds significant overheads,  vulnerable to shift-and-leak~\cite{shift-and-leak-iccad19} attack and increases test generation effort. {\em EFF}~\cite{rajit:encryptFF} addresses these issues by locking scan flip-flip outputs (FOs). 
But {\em EFF} is insecure against {\em ScanSAT}~\cite{lilas:aspdac19}, thus there is a need for a better defense that is both secure and practical.

\section{Preliminaries}
\label{sec:pre}

This section discusses the vulnerability of prior work on scan-locking and highlights the threat model.

\vspace{-0.15em}
\subsection{Scan-Locking$\!$\cite{rajit:encryptFF} \& State-of-the-Art Attacks~\cite{lilas:aspdac19,pramod-faa, smt-attack}}
\label{sec:eff}
\noindent 
In {\em EFF} technique~\cite{rajit:encryptFF}, flip-flops (FFs) on the non-critical timing paths of a sequential circuit are selected, and XOR/XNOR-type key gates are added to lock the $Q$-outputs, which drive combinational logic as well as the scan-chain. Figure~\ref{fig:eff-seql-scansat}(a) shows a sample sequential circuit with $2$ out of $4$  FFs encrypted using {\em EFF}-style scan-locking. In Figure~\ref{fig:eff-seql-scansat}(a):
\begin{itemize}
    \item $G_0$ and $G_1$ are the primary inputs. $SI$ and $SO$ are the circuit's scan-input port and scan-output port respectively; 
    \item FFs $1$ and $2$ have feedback, while FFs $3$ and $4$ do not have feedback. $G_2$, $G_4$, $G_6$ and $G_8$ are corresponding FF inputs (FIs) respectively. $G_3$, $G_5$, $G_7$ and $G_9$ are corresponding FF outputs (FOs) respectively; and 
    \item $ck_0$ and $ck_1$ are the combinational key bits, while $fok_0$ and $fok_1$ are key bits used to lock the FO $G_3$ (XOR-type key gate) and FO $G_5$ (XNOR-type key gate) respectively.
\end{itemize}

ScanSAT~\cite{lilas:aspdac19} shows that it is possible to convert this scan-locked instance to the scan-unrolled locked instance of Figure~\ref{fig:eff-seql-scansat}(b), launch the SAT-attack on the unrolled instance and decrypt the functionally correct sequential key. Here, in scan-mode of operation:
$SI(G_3)$ and $SI(G_5)$ are the scan-input-bits corresponding to FFs $1$ and $2$ respectively; and $ESO(G_2)$ and $ESO(G_4)$ are the locked-scan-output bits corresponding to FFs $1$ and $2$ respectively.
Hence, {\em EFF} technique is not secure. 
Similar to {\em ScanSAT}~\cite{lilas:aspdac19}, it is possible to extend some of the state-of-the-art attacks like {\em HackTest}-attack~\cite{hacktest},  functional-analysis-attacks on logic-locking (FALL)~\cite{pramod-faa}, and SMT-attack~\cite{smt-attack}. We thus evaluate {\em SeqL} on all these attacks.

\subsection{Threat model}
\label{sec:attack-model}

\noindent We consider a malicious foundry that offers fabrication, assembly and testing services~\cite{tsmc}. 
Thus, the attacker has access to layout and mask information, and is thus able to reverse-engineer the gate-level netlist.  
There are two possible scenarios: (1) The attacker uses an activated IC ({\it oracle}), and applies scan patterns to the IP, and observes corresponding scan responses in EDT-bypass mode. Since the IP is located somewhere deep inside the {\em SoC}, we assume that the IP is {\bf controllable/observable, only through scan-chains}. Typically, scan ports are not deactivated to facilitate debug of customer returns, which the attacker exploits to launch the SAT-attack; or (2) The attacker is at the outsourced tester, where the attacker can place the dies in EDT-bypass mode, applies desired scan patterns to the IP, and observes corresponding scan responses.



\section{Solution Insight}
\label{sec:insight}

\noindent As discussed in the previous section, when SAT-attack is launched on the scan-unrolled {\em EFF}-style scan-locked circuit, the SAT solver returns the functionally correct key.
In the discussion that follows, we exploit the following principles:
\begin{enumerate}
    \item In {\em EFF}-style scan-locking, the FO key-gate corresponding to the FF appears {\em both} in the scan-input and scan-output paths of the FF in the scan-unrolled instance;
    \item Functional path can be isolated from the locked scan path used by the attacker, thus achieving resilience; 
    \item To prevent vulnerability to multi-cycle-attack~\cite{guin:2018}, it is important to selectively lock FFs without feedback; 
    \item Since the FO key gates cascade with FI key gates to form XOR/XNOR-chains, it is possible to obfuscate the solver. 
\end{enumerate}

Figure~\ref{fig:eff-seql-scansat}(c) shows the proposed {\em SeqL}-style scan-locking by transforming the circuit in Figure~\ref{fig:eff-seql-scansat}(a), using above principles. Figure~\ref{fig:eff-seql-scansat}(c) is different from Figure~\ref{fig:eff-seql-scansat}(a) in the following ways:
\begin{itemize}
    \item There is a separate $Q$ and $SQ$, and the key gate is added at $SQ$ ($SQ$ key-gate), thus leaving the functional output $Q$ unencrypted. This is referred to as {\em functional isolation};
    \item FFs without feedback e.g., $3$ and $4$ are selected for locking; 
    \item $sqk_0$ and $sqk_1$ are the key bits used to lock the $SQ$ output of FFs $3$ (with XOR-type key gate) and $4$ (with XNOR-type key gate) respectively; and 
    \item Extra key gates (both of XOR type and without additional obfuscation logic in this case, for ease of explanation) are added to lock FIs of both these FFs, using $fik_0$ and $fik_1$ key bits respectively. These key gates are referred to as FI key gates in the rest of this paper.  
\end{itemize}
Figure~\ref{fig:eff-seql-scansat}(d) shows the corresponding scan-unrolled equivalent combinational circuit. 
{\it The purple dashed line is the functional boundary. This means that the key gates to the right of this boundary ($SQ$ key-gates) only affect scan-operation, and do not affect normal functional operation of the circuit.} This is because the attacker uses the scan mode of operation, and hence observes $ESO(G_2)$ and $ESO(G_4)$. 
However, the circuit's normal functional operation is purely influenced by $E(G_2)$ and $E(G_4)$, and hence the XOR/XNOR-chains (in red) cease to exist. This renders the scan-correct decrypted key, being {\em functionally incorrect}. 
After running SAT-attack on circuit in Figure~\ref{fig:eff-seql-scansat}(d), when the combinational portions of the sequential circuit, the original unencrypted sequential circuit and the solver key are inputted to the formal equivalence checker, the result was {\em not equivalent}. Hence, by functional isolation and FI locking,  resilience was achieved, thus making the proposed {\em SeqL}-style scan-locking mechanism secure. Next subsection explains this behavior using an abstract model.

\begin{figure}[!t]
  \begin{center}
    \subfigure[{\em EFF}-style ]{\label{fig:s27-EFF-ppof}\includegraphics[scale=0.4]{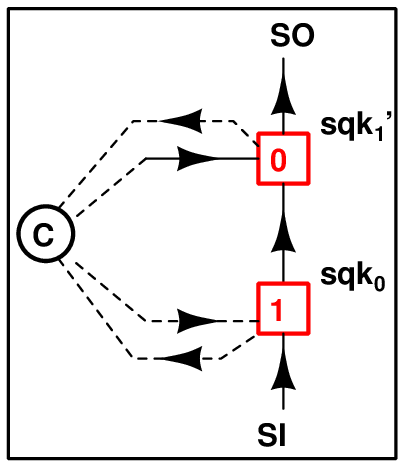}}
    \subfigure[{\em SeqL}-style ]{\label{fig:s27-SeqL-ppof}\includegraphics[scale=0.4]{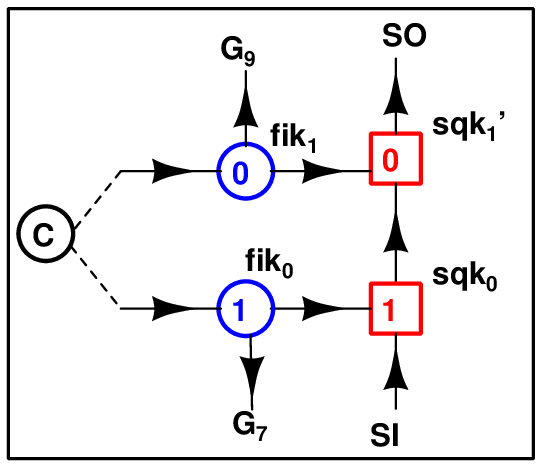}}
  \end{center}
    \caption{Abstract models for circuits in Figures~\ref{fig:eff-seql-scansat}(a) and~\ref{fig:eff-seql-scansat}(c)}
  \label{fig:ppof}
\end{figure}

\subsection{Abstract model}

\label{sec:ppof}
\indent Figure~\ref{fig:ppof} shows an abstracted model of the sequential circuit, with the combinational logic abstracted into a source-sink circular vertex {\em C}, each FI key-gate abstracted into a blue circular vertex, and each FF key-gate abstracted into a red rectangular vertex. Figures~\ref{fig:s27-EFF-ppof} and~\ref{fig:s27-SeqL-ppof}  show models corresponding to circuits in Figures~\ref{fig:eff-seql-scansat}(a) and~\ref{fig:eff-seql-scansat}(c) respectively. 

In the abstract model corresponding to the proposed scan-locking shown in Figure~\ref{fig:s27-SeqL-ppof}, the following are functional paths:
\begin{itemize}
    \item $FP_0: fik_0 \longrightarrow G_7$
    \item $FP_1: fik_1 \longrightarrow G_9$
\end{itemize}
and the following are scan-out paths:
\begin{itemize}
    \item $SP_0: fik_0 \longrightarrow sqk_0 \longrightarrow sqk_1^{'} \longrightarrow SO$
    \item $SP_1: fik_1 \longrightarrow sqk_1^{'} \longrightarrow SO$
\end{itemize}
Figure~\ref{fig:s27-SeqL-ppof} shows that the number of inversions for the scan-output-paths $SP_0$ and $SP_1$, are $2$ and $0$, respectively. Since all scan-output-paths have even inversion parity, the proposed locked circuit is {\em correct for scan operation}.
However, for functional paths, the number of inversions for $FP_0$ and $FP_1$, are $1$ and $0$. Since functional path $FP_0$ has odd-inversion-parity, the circuit is {\em incorrect for functional operation}.

\begin{figure}[!t]
\centering
\includegraphics[scale=0.54]{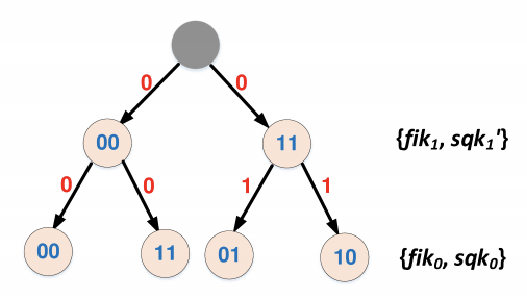}
\caption{Key Assignment Graph (KAG) for  circuit in Figure~\ref{fig:eff-seql-scansat}(c). $KAG$ is a binary tree with dummy root  node, the leaves of which correspond to the rows in Table~\ref{tab:truth-table-sequential-lock} whose scan-correctness column is \color{blue}{TRUE}.}
\label{fig:sample-circuit-seql-locked-kag}
\end{figure}

To understand the behavior systematically, Table~\ref{tab:truth-table-sequential-lock} enumerates all possibilities for the scan-lock $\{fik_1, sqk_1^{'}, fik_0, sqk_0\}$ for the circuit in Figure~\ref{fig:eff-seql-scansat}(c). 
\begin{table}[!t]
\begin{center}
\scriptsize
\caption{Truth table of our proposed  scan-lock in Figure~\ref{fig:eff-seql-scansat}(c)}
\label{tab:truth-table-sequential-lock}
\begin{tabular}{|c|c|c|c|c|c|}
\hline
$fik_1$ & $sqk_1'$  & $fik_0$ & $sqk_0$ & Scan-Correct & Functional-Correct\\
\hline
 $0$ & $0$ & $0$ & $0$ & \color{blue}{TRUE} &  \color{blue}{TRUE}  \\
\hline 
$0$ & $0$ & $0$ & $1$ & \color{red}{FALSE} &  \color{blue}{TRUE}  \\
\hline 
 $0$ & $0$ & $1$ & $0$ & \color{red}{FALSE} &  \color{red}{FALSE}  \\
\hline 
 $0$ & $0$ & $1$ & $1$ & \color{blue}{TRUE} &  \color{red}{FALSE}  \\
 \hline
 $0$ & $1$ & $0$ & $0$ & \color{red}{FALSE} &  \color{blue}{TRUE} \\
\hline 
 $0$ & $1$ & $0$ & $1$ & \color{red}{FALSE} &  \color{blue}{TRUE}  \\
\hline 
 $0$ & $1$ & $1$ & $0$ & \color{red}{FALSE} &  \color{red}{FALSE}  \\
\hline 
 $0$ & $1$ & $1$ & $1$ & \color{red}{FALSE} &  \color{red}{FALSE}  \\
 \hline
 $1$ & $0$ & $0$ & $0$ & \color{red}{FALSE} &  \color{red}{FALSE}  \\
\hline 
 $1$ & $0$ & $0$ & $1$ & \color{red}{FALSE} &  \color{red}{FALSE}  \\
\hline 
 $1$ & $0$ & $1$ & $0$ & \color{red}{FALSE} &  \color{red}{FALSE}  \\
\hline 
 $1$ & $0$ & $1$ & $1$ & \color{red}{FALSE} &  \color{red}{FALSE}  \\
 \hline
 $1$ & $1$ & $0$ & $0$ & \color{red}{FALSE} &  \color{red}{FALSE}  \\
\hline 
 $1$ & $1$ & $0$ & $1$ & \color{blue}{TRUE} &  \color{red}{FALSE}  \\
\hline 
 $1$ & $1$ & $1$ & $0$ & \color{blue}{TRUE} &  \color{red}{FALSE}  \\
\hline 
 $1$ & $1$ & $1$ & $1$ & \color{red}{FALSE} &  \color{red}{FALSE}  \\
 \hline
\end{tabular}
\end{center}
\end{table}
The rows in this table, that show up as $TRUE$ for the scan-correct column, are the possible keys returned by the SAT-solver. Among the four functionally-correct rows only one is scan-correct, thus {\em SeqL} {\bf hides} $\frac{3}{4}$ functional-correct keys from the attacker. Similarly, among the four scan-correct rows, \emph{only one} is functionally correct, thus {\em SeqL} maximizes the odds against the functionally-correct-key among the scan-correct-keys.
Figure~\ref{fig:sample-circuit-seql-locked-kag} shows the corresponding key assignment graph (KAG).
The sequential key returned by the solver is the second leaf from the left. Since this leaf is a functionally incorrect key, the technique is able to achieve resilience. In this example, odds against the functionally correct key is $p = \frac{3}{4} = 0.75$.


\subsection{Analysis}
\label{sec:analysis}
This section formally analyzes the security of logic locking and proves that if {\em SeqL} is used to lock $n$ flip-flops in the sequential circuit, then the odds against the functionally-correct-key among the scan-correct-keys equals $  1-\frac{1}{2^{n}}$, assuming the attack is launched in EDT-bypass mode. 

Given an FI-SQ key-pair $\{fik_i, sqk_i \}$, there are 4 possible assignments $\{00, 01, 10, 11\}$. Let $n$ be the number of locked FI-SQ pairs. Let $KAG=(V,E)$ be a vertex-labelled edge-weighted directed graph, where the vertices correspond to FI-SQ pairs and the edges correspond to inversion parity. The direction of edges is opposite to the scan-out-path direction. In $KAG$, the children of every vertex at depth $i$ from the root correspond to $i^{th}$ flip-flop from the end of the scan-out-path. All node and edge assignments are performed to ensure scan-correctness. $KAG$ is a  tree, whose root vertex is a dummy node, with exactly two children $00$ and $11$. 

The labels on the vertices in $KAG$ are $00, 01, 10$ or $11$, corresponding to $\{fik_i, sqk_i\}$, $\{fik_i, sqk_i^{'}\}$, $\{fik_i^{'}, sqk_i\}$ or $\{fik_i^{'}, sqk_i^{'}\}$ depending on whether FI key-gate, SQ key-gate combination is $\{$XOR, XOR$\}$, $\{$XOR, XNOR$\}$, $\{$XNOR, XOR$\}$ or $\{$XNOR, XNOR$\}$ respectively. $00$ and $11$ are even-parity vertices, whereas $01$ and $10$ are odd-parity vertices. The children of $00$ and $01$ are even-parity vertices. The children of $10$ and $11$ are odd-parity vertices. Hence, every non-root vertex has exactly 2 children. The possible weights on the edges in $KAG$ are $0$ or $1$, which signifies parity. The parity of an edge signifies the presence/absence of signal-inversion at the child flip-flop, which is same as the parity of the corresponding child vertex. $inv_k$ equals 0 or 1, depending on whether $k^{th}$ flip-flop along the scan-chain from the scan-output is locked with an $XOR$ or $XNOR$ key-gate respectively. \\


\noindent {\em \underline{Theorem 1}:} Parities of both edges of a vertex are identical. \\
{\em Proof:} Assume vertex $v_i$ in $KAG$ at depth $i$. In order to ensure scan-correctness, ($fik_i \oplus sqk_i \oplus inv_i) \oplus \sum_{k=1}^{i-1} (sqk_k \oplus inv_k )$ should equal $0$. If $\sum_{k=1}^{i-1} (sqk_k \oplus inv_k)$ equals 0, $( fik_i \oplus sqk_i \oplus inv_i)$ becomes 0 (possible children of $v_i$ are 00 and 11, in both cases parity of edge is 0).\\
On the other hand, if $\sum_{k=1}^{i-1} (sqk_k \oplus inv_k)$ equals 1, $( fik_i \oplus sqk_i \oplus inv_i)$ becomes 1 (possible children of $v_i$ are 01 and 10, in both cases parity of edge is 1).
Thus, parity of left and right edges of a vertex are identical, hence the proof. \\

\noindent {\em \underline{Theorem 2}:} $KAG$ is a binary tree. \\
{\em Proof:} Root vertex has exactly two children. Additionally, every non-root vertex has exactly two children. Since every vertex has exactly two children, $KAG$ is a binary tree, hence the proof. \\


\noindent {\em \underline{Theorem 3}}: The odds against the functionally-correct-key among the scan-correct-keys is $p = 1-\frac{1}{2^{n}}$\\
{\em Proof:} The path from root to a functionally correct leaf should have all $00$ nodes, there is exactly one such leaf in $KAG$. Since, the total number of leaves in $KAG$ $=2^n$, odds $p = \frac{2^n - 1 }{2^n} =  1-\frac{1}{2^{n}}$, hence the proof. \\



\begin{figure}[!t]
\centering
\includegraphics[scale=0.61,center]{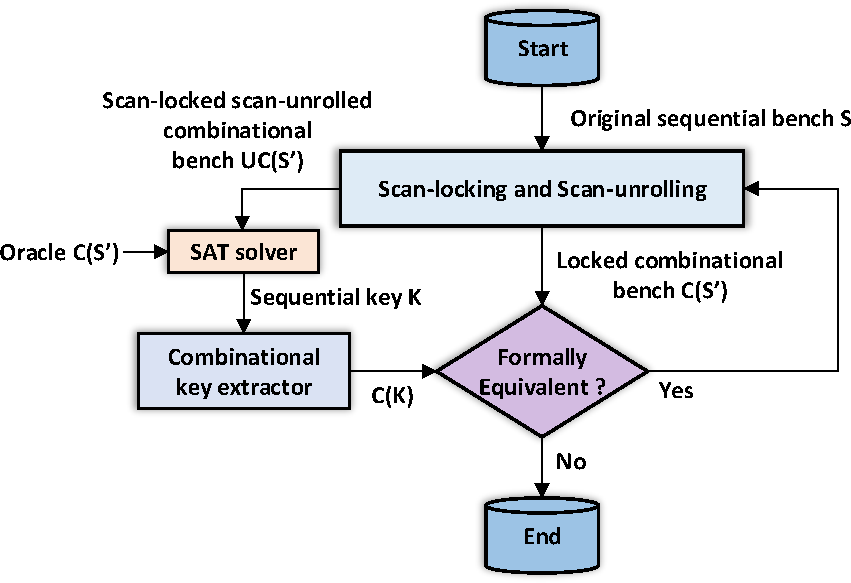}
\caption{Automating SeqL}
\label{fig:flowchart}
\end{figure}

\section{Automating S\lowercase{eq}L Defense}
\label{sec:proposed-scheme}
So far, we have seen the effectiveness of {\em SeqL} in defending attacks on scan locking. This section shows how to automate {\em SeqL}, so that it can be practically deployed on large circuits.\\

\noindent {\bf Objective:} Lock selective scan flip-flops (FI-SQ pairs) without feedback such that functional output corruption is achieved, while area-overhead is minimized. \\


\noindent {\bf Solution:} The likelihood of functional output corruption is maximized with increase in $p=1-\frac{1}{2^{n}}$, whereas area-overhead increases linearly with $n$. Hence, the chances of functional output corruption increases rapidly with $n$, with minimal increase in area overhead.  Figure~\ref{fig:flowchart} shows \emph{SeqL} flow, which exploits this principle to iteratively lock scan flip-flops (FI-SQ pairs) from the end of the scan-chain(s) until functional output is corrupted. Thus, the proposed scan-locking solution has two parts:
\begin{enumerate}
\item An functionally-isolated scan-locked flip-flop design.
\item An iterative FI-SQ locking algorithm.
\end{enumerate}


\begin{figure}[!t]
\centering
\hspace*{-1em}
\includegraphics[scale=0.2,center]{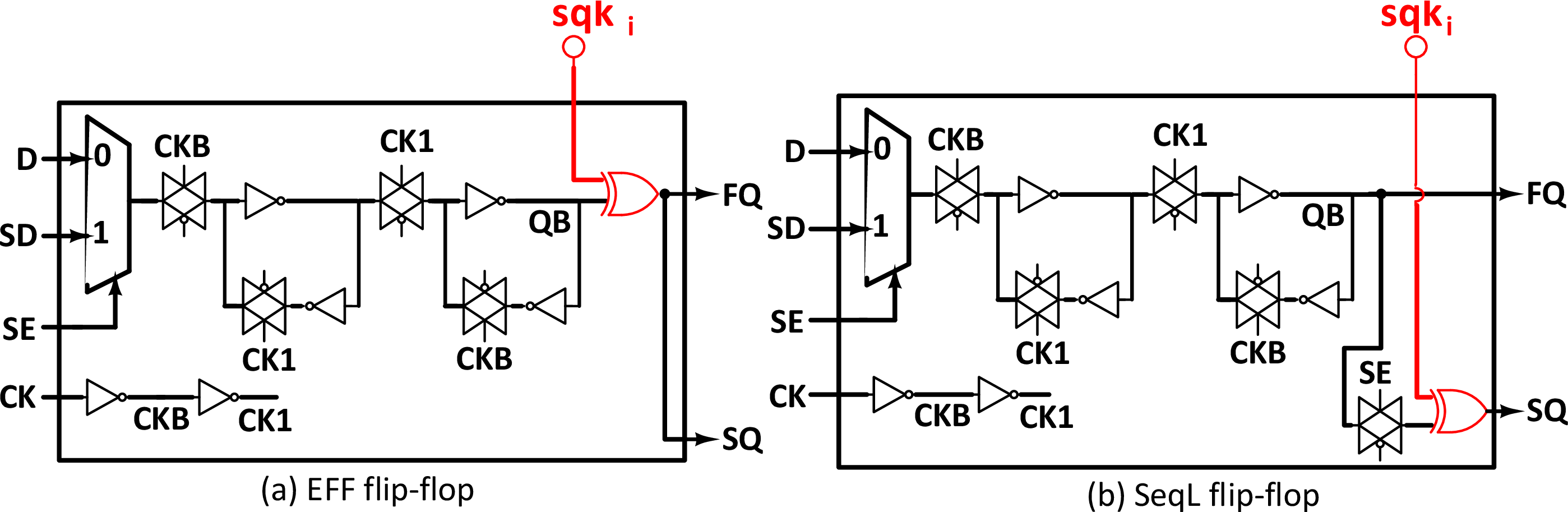}
\caption{Flip-flop variants for scan-locking}
\label{fig:ffs}
\end{figure}

\subsection{Functionally-isolated scan-locked flip-flop design}
\label{sec:isolation-based-ff-locking}
\noindent We define the sequential key to be $K=\{K_c, K_{fi}, K_{sq}\}$, where $K_c$, $K_{fi}$ and $K_{sq}$ are portions of the key that lock the combinational logic (excluding the FIs), the FIs and the SQs respectively. In {\em EFF} technique, all these components influence the sequential circuit's normal functional operation. Figure~\ref{fig:ffs}(a) shows the {\em EFF}-style scan-locking scheme, where the $FO$ key gate output, is broadcasted to {\em Scan-Q} (referred to as $SQ$ in the figure), as well as {\em Functional-Q} (referred to as $FQ$ in the figure). The proposed isolation-based scan-locking is shown in  Figure~\ref{fig:ffs}(b), which isolates the functional path from the locked scan path. Hence, the $SQ$ key gate locks only $SQ$ and has no influence on $FQ$. Thus, in the proposed {\em SeqL} technique, only  $K_c$ and $K_{fi}$ influence (while $K_{sq}$ has no effect on) the sequential circuit's normal functional operation. This assists in returning the {\em functionally incorrect key}, thus aiding in functional output corruption when applied with the key returned by the SAT-solver. There is an additional transmission gate added to this structure in the scan path to avoid toggling of the locking key gate along the scan path. Although this adds 2 extra transistors per flip-flop, the overhead is marginal compared to the benefit of savings obtained in Energy-Per-Toggle ($EPT$) of the flip-flop during normal functional operation, similar to CSP-scan~\cite{Potluri:2015}. The comparison of area, timing and $EPT$ of the {\em EFF} as well as {SeqL} flip-flops are provided later in Section~\ref{sec:results}. 

\subsection{Iterative key pushing algorithm (IKPA) for pipelined combinational circuits}
\label{sec:algorithm}
\noindent 
Algorithm~\ref{alg:ipla-c-pipelined} shows the iterative key gate pushing algorithm, that takes a logic-locked combinational circuit with pipeline stages both at its inputs and outputs. 
Since the circuit already has key gate overhead, to avoid any further overhead, the algorithm iteratively pushes some of the key gates inside combinational logic to the output boundary. 
\begin{algorithm}[!t]

\title{Iterative key pushing algorithm for pipelined logic-locked combinational circuits}
\caption{Iterative key pushing algorithm for pipelined logic-locked combinational circuits}
\label{alg:ipla-c-pipelined}
\KwIn{$C$}

\BlankLine
\While{$C^{'} = C$}{
    Identify a combinational key gate pair $k_c$, an unvisited FI-SQ pair {\it for a flip-flop without feedback}  $k_b$ and mark corresponding $k_b$ as visited \;
    Push $k_c$ to $k_b$\;
    Run SAT-solver and update $K_{fi}$, $C^{'}$\;
}
\KwResult{$C^{'}$, $K_{fi}$}
\end{algorithm}
We measure the success of the {\em IKPA} algorithm using formal equivalence checking. If the returned solver key makes the two circuits {\em different} or in other words {\em not equivalent}, then we are successful i.e., {\em functional output corruption is achieved}. Additionally, even with {\em SeqL}, we have found that in all the cases, the decrypted $K_c$ is correct. Hence, it is only $K_{fi}$, that causes functional output corruption. 
Section~\ref{sec:results} demonstrates the detailed results. 

\subsection{Iterative boundary locking algorithm (IBLA) for sequential circuits}
\label{sec:ibla}
\noindent Unlike pipelined logic-locked combinational circuits, there are no existing key gates for sequential circuits. Hence, FI-SQ locking or in other words, boundary locking has to be done afresh. Moreover, since $K_c$ is always successfully decrypted, combinational logic except FIs is not locked to ensure cost-effectiveness. 
Since the higher the number of inserted key gates at the FI-SQ boundary, the higher the area overhead, we make this parameter, $\gamma$, user-configurable. 
\setlength{\textfloatsep}{8pt}
\begin{algorithm}[!t]

\title{IBLA for sequential circuits}
\caption{Iterative boundary locking algorithm for sequential circuits}
\label{alg:ibla}
\KwIn{$S$, $K_{sq}$}

\BlankLine
\While{$S^{'} = S$ and $|K_{fi}| + |K_{sq}| <= \gamma$}{
    Identify an unvisited FI-SQ pair {\it for a FF without feedback} and mark corresponding pair as visited \;
    Obfuscate corresponding FI-SQ pair\;
    Run SAT-solver and update $K_{fi}$, $S^{'}$\;
}
\KwResult{$S^{'}$, $K_{fi}$}
\end{algorithm}
Algorithm~\ref{alg:ibla} takes a sequential benchmark as input, and iteratively obfuscates unvisited FI-SQ pairs with XOR/XNOR-type key-gates, until functional output corruption is achieved or $|K_{fi}| + |K_{sq}| > \gamma$. 
Since the implementation is simple and security is achieved with low overheads, this is attractive for industry practice.

\vspace{-1.0em}
\section{Experimental Evaluation}
\vspace{-0.75em}
\label{sec:results}

\noindent We validate the security of {\em SeqL} against a multitude of state-of-the-art attacks and quantify its reduced overheads compared to prior work. This analysis confirms our claims on genericness, robustness, and scalability of {\em SeqL}.
Algorithms~\ref{alg:ibla} ({\em IBLA}) and~\ref{alg:ipla-c-pipelined} ({\em IKPA}) were used for scan-locking the sequential benchmarks and pipelined combinational benchmarks, respectively. Both the locking algorithms were implemented in {\em Perl}. Since the locking algorithm execution times across all the benchmarks were very small (matter of seconds), the execution times were not reported.

\begin{table}[!t]
\renewcommand{\arraystretch}{0.78}
\setlength\extrarowheight{2pt}
\setlength{\tabcolsep}{5pt}
\begin{center}
\vspace{-1em}
\caption{Resilience of {\em SeqL} for Pipelined Combinational Benchmarks for $5\%$ logic locking. {\em '\bluecheck' $ $ is secure and '\redx' $ $ is insecure.}}
\vspace{-.5em}
\label{tab:results-proposed-eff-c}
\begin{tabular}{|c|c|c|c|c|c|c|c|c|}
\hline
Bench. & \multicolumn{2}{c|}{RND} & \multicolumn{2}{c|}{DAC'12} & \multicolumn{2}{c|}{ToC'13/xor} & \multicolumn{2}{c|}{ToC'13/mux} \\
\hline
	& {\em EFF}  & {\em SeqL} &
	{\em EFF} & {\em SeqL} & 
	{\em EFF} & {\em SeqL} & 
	{\em EFF} & {\em SeqL}  \\

\hline 
\texttt{apex2}      
&\redx	&\bluecheck		&\redx &\bluecheck			
&\bluecheck &\bluecheck		&\redx &\bluecheck \\

\hline  
\texttt{apex4}   
&\redx	&\bluecheck			&\redx &\bluecheck		
&\bluecheck &\bluecheck		&\bluecheck &\bluecheck \\

\hline 
\texttt{i4}       
&\bluecheck &\bluecheck 		&\redx &\bluecheck		
&\bluecheck &\bluecheck		&\redx &\bluecheck \\

\hline 
\texttt{i7}      
&\bluecheck &\bluecheck 		&\redx &\bluecheck		
&\bluecheck &\bluecheck		&\redx &\bluecheck \\

\hline 
\texttt{i8}     	&\bluecheck &\bluecheck		&\redx &\bluecheck		
&\bluecheck &\bluecheck		&\redx &\bluecheck \\

\hline 
\texttt{i9}     	&\bluecheck &\bluecheck		&\redx &\bluecheck		
&\redx &\bluecheck				&\redx &\bluecheck \\

\hline 
\texttt{seq}     
&\bluecheck & \bluecheck		&\redx &\bluecheck		
&\bluecheck &\bluecheck		&\redx &\bluecheck \\

\hline 
\texttt{k2}     	&\bluecheck &\bluecheck		&\redx &\bluecheck		
&\bluecheck &\bluecheck		&\redx &\bluecheck \\


\hline 
\texttt{ex1010}
&\bluecheck &\bluecheck 		&\redx &\bluecheck		
&\bluecheck &\bluecheck		&\redx &\bluecheck \\

\hline
\texttt{dalu}    
&\bluecheck	&\bluecheck	&\redx &\bluecheck		
&\bluecheck &\bluecheck		&\redx &\bluecheck \\

\hline 
\texttt{des}     
&\bluecheck &\bluecheck 		&\redx &\bluecheck	
&\bluecheck &\bluecheck		&\redx &\bluecheck \\

\hline 
\texttt{c432}   
&\redx &\bluecheck			&\redx &\bluecheck		
&\redx &\bluecheck 			&\redx &\bluecheck  \\

\hline 
\texttt{c499}    
&\redx &\bluecheck 			&\redx &\bluecheck 		
&\redx &\bluecheck				&\redx &\bluecheck \\

\hline 
\texttt{c880}    
&\bluecheck	&\bluecheck	&\bluecheck &\bluecheck	
&\redx &\bluecheck			&\redx &\bluecheck \\

\hline 
\texttt{c1355}   
&\bluecheck &\bluecheck		&\redx &\bluecheck		
&\bluecheck &\bluecheck		&\redx &\bluecheck \\

\hline
\texttt{c1908}   
&\bluecheck &\bluecheck		&\redx &\bluecheck		
&\redx &\bluecheck 			&\redx &\bluecheck \\

\hline 
\texttt{c3540}   &\bluecheck &\bluecheck 		&\redx &\bluecheck		&\redx 
&\bluecheck				&\redx &\bluecheck \\

\hline 
\texttt{c5315}   &\bluecheck &\bluecheck		&\bluecheck &\bluecheck	&\redx 
&\bluecheck			&\redx &\bluecheck \\

\hline 
\texttt{c7552}   &\bluecheck &\bluecheck		&\redx &\bluecheck		&\redx 
&\bluecheck			&\redx &\bluecheck \\

\hline 
\end{tabular}
\end{center}
\vspace{-1.25em}
\end{table}

\vspace{-.5em}
\subsection{Resilience of SeqL vs. EFF~\cite{rajit:encryptFF} against SAT-Attacks on pipelined combinational benchmarks}
\vspace{-.25em}
Table~\ref{tab:results-proposed-eff-c} shows the results of applying the procedure shown earlier in Figure~\ref{fig:flowchart} on 4 different encryption schemes validated in~\cite{pramod:host15}, and compared against {\em EFF}~\cite{rajit:encryptFF}. 
This table shows that {\em SeqL} secured all sequential circuits against SAT-attack in $100 \%$ of the cases.  As explained in Section~\ref{sec:insight}, (1) $K_c$ was successfully decrypted in all cases, while (2) $K_{fi}$ was incorrect, hence causing functional output corruption, thus achieving resilience.
Results on $IOLTS'14$ encryption scheme~\cite{dupuis:iolts14, pramod:host15} gave 0\% resilience in {\em EFF} case and 100\% resilience in {\em SeqL} case, across all benchmarks, hence not reported in Table~\ref{tab:results-proposed-eff-c} for brevity. 

\begin{table}[!t]
\begin{threeparttable}
\renewcommand{\arraystretch}{0.78}
\setlength\extrarowheight{2pt}
\centering
\caption{Resilience of {\em SeqL} against state-of-the-art attacks on pipelined combinational benchmarks.\vspace{-0.75em}} 
\label{tab:results-sota-attacks}
\begin{tabular}{|c|c|c|c|c|c|}
\hline
       & \multicolumn{3}{|c|}{Oracle-guided} & \multicolumn{2}{|c|}{Oracle-less} \\
\hline
Bench. & DDIP~\cite{double-dip} & SS~\cite{lilas:aspdac19} & SMT~\cite{smt-attack} & HT~\cite{hacktest} & FALL~\cite{pramod-faa}\\

\hline 
\texttt{apex2}      
&\bluecheck  &\bluecheck &\bluecheck &\bluecheck &\bluecheck \\

\hline  
\texttt{apex4}   
&\bluecheck  &\bluecheck &\bluecheck  &\bluecheck & \color{red}{{\bf NK}}\\

\hline 
\texttt{i4}       
&\bluecheck  &\bluecheck &\bluecheck &\bluecheck &\bluecheck \\

\hline 
\texttt{i7}      
&\bluecheck  &\bluecheck &\bluecheck &\bluecheck &\bluecheck \\

\hline 
\texttt{i8}
&\bluecheck  &\bluecheck &\bluecheck &\bluecheck &\bluecheck \\

\hline 
\texttt{i9}     	
&\bluecheck  &\bluecheck &\bluecheck &\bluecheck &\bluecheck \\

\hline 
\texttt{seq}     
&\bluecheck  &$-$ &\bluecheck  &$-$  &\bluecheck \\

\hline 
\texttt{k2}     
&$-$ &\bluecheck &$-$ &\bluecheck &\bluecheck \\


\hline 
\texttt{ex1010}  
&\bluecheck  &\bluecheck &\bluecheck &\bluecheck & \color{red}{{\bf NK}} \\

\hline
\texttt{dalu}    
&\bluecheck  &\bluecheck &\bluecheck &\bluecheck &\bluecheck \\

\hline 
\texttt{des}     
&$-$  &\bluecheck &\color{red}{{\bf NK}} &\bluecheck &\bluecheck\\

\hline 
\texttt{c432}   
&\bluecheck  &\bluecheck &\bluecheck &\bluecheck &\bluecheck  \\

\hline 
\texttt{c499}    
&\bluecheck &\bluecheck &\bluecheck &\bluecheck &\bluecheck \\

\hline 
\texttt{c880}    
&\bluecheck  &\bluecheck &\bluecheck &\bluecheck &\bluecheck \\

\hline 
\texttt{c1355}   
&\bluecheck  &\bluecheck &\bluecheck &\bluecheck &\bluecheck \\

\hline
\texttt{c1908}   
&\bluecheck &\bluecheck &\bluecheck &\bluecheck &\bluecheck \\

\hline 
\texttt{c3540}   
&\bluecheck  &\bluecheck &\bluecheck &\bluecheck &\bluecheck \\

\hline 
\texttt{c5315}   
&$-$  &\bluecheck &\bluecheck &\bluecheck &\bluecheck\\
    
\hline 
\texttt{c7552}   
&\color{red}{{\bf NK}}  &\bluecheck &\color{red}{{\bf NK}} &\bluecheck &\bluecheck \\

\hline 
\end{tabular}
\begin{tablenotes}
\item All  experiments are run on {\em IBM BladeCenter$\textsuperscript{\textregistered}$ Cluster} with abort-limit of 1 week. {\em '\bluecheck' $ $ is secure and '\redx' $ $ is insecure. '-' indicates decryption time exceeds abort-limit, while 'NK' indicates \texttt{No-Key}.}
\end{tablenotes}
\end{threeparttable}
\vspace{-1em}
\end{table}

\begin{table*}[!t]
\centering
\renewcommand{\arraystretch}{0.78}
\setlength\extrarowheight{2.2pt}
\vspace{-1em}
\setlength{\tabcolsep}{3.5pt}
\begin{threeparttable}[t]
\centering
\caption{Resilience of {\em SeqL} for Sequential Circuits. The Scan-locking was done using {\em IBLA} algorithm.\vspace{-.75em}}
\label{tab:results-proposed-eff-sequential}
\begin{tabular}{|c|c|c|c|c|c|c|c|c|c|c|c|c|c|c|c|c|c|c|c|c|}
\hline
Bench.	  & \#{\em Gates} & \#{\em SFFs} & \#{\em SCs} & $|R_{wof}|$ & \multicolumn{2}{c|}{{\em EFF}~\cite{rajit:encryptFF}} & \multicolumn{9}{|c|}{{\em SeqL}} \\

\hline
	  & & & & &  &  &  & &  & \multicolumn{3}{c|}{Resilience} & \multicolumn{3}{c|}{Decryption Time} \\
\hline
	&  & & & &  Res. & Ov. 			  & $n$ & $p$& Ov.  &   N=1 & N=2 & N=5 & N=1 & N=2 & N=5\\
\hline
\texttt{b14} & 10,012 &245 &3 & $54$ & \redx & $3.3$\%    & 8 & $0.99$& $0.24$ \%   & \bluecheck &\bluecheck  &\bluecheck   &$19\ min$  & $2\ min$  & $2\ min$\\
\hline
\texttt{b15} & 12,992 &449 &5 & $70$ & \redx &$4.3$ \%    & 9 & $0.99$& $0.2$ \%   & \bluecheck &\bluecheck  &\bluecheck  &$47\ min$  & $11\ min$   &$ 164\ min$ \\
\hline
\texttt{b17} & 32,192 &1,415 &15 & $97$ & \redx &$5.2$ \%   & 6 & $0.99$& $0.05$ \%   & \bluecheck &\bluecheck  &\bluecheck  &$10\ min$  &$17\  hrs.$    & $47\  hrs.$\\
\hline
\texttt{b18} & 114561 & 3,320 & 34 & $23$ & \redx  & $3.8$ \%   &10 &$0.99$ & $0.03$ \% &\bluecheck &  $-$ &$-$  &$53\  hrs.$  &  \color{red}{$>$ {\bf abort-limit}}  &\color{red}{$>$ {\bf abort-limit}} \\
\hline
\texttt{b19} & 231,266 & 6,642 & 67 & $30$ & \redx  & $3.7$ \%   &10 &$0.99$ &  $0.01$ \% &\bluecheck  & $-$ & $-$ &$91\  hrs.$  &\color{red}{$>$ {\bf abort-limit}}    &\color{red}{$>$ {\bf abort-limit}}\\
\hline
\texttt{b20} &20,172 &490 &5  & $22$ & \redx  & $3.3$ \%   &10 &$0.99$ & $0.15$ \%   &\bluecheck &\bluecheck  &\bluecheck  &$7\ min.$  &$15\ min.$    &$37\ min.$\\
\hline
\texttt{b21} &20,517 &490 &5  & $22$  & \redx  & $3.2$ \%   &10 &$0.99$ &$0.15$ \%   &\bluecheck &\bluecheck  &\bluecheck  &$6\ min.$  &$34\ min.$    &$36\ min.$\\
\hline
\texttt{b22} &29,897 &735 &8 & $22$ & \redx  & $3.3$ \%   &10 &$0.99$ &$0.1$ \%   &\bluecheck &\bluecheck  &\bluecheck  &$11\ min.$  &$37\ min.$    &$67\ min$\\
\hline 
\texttt{RISC-V flat.} & 25,096 & 2,031 & 20 & $226$ & \redx& $7.9$ \% &10 &$0.99$ & $0.09$ \% &\bluecheck &\bluecheck  &\bluecheck &$2\ min.$ &$13\ min.$ &$6\  hrs.$ \\
\hline
\end{tabular}
\begin{tablenotes}
\item[1]  $N$ denotes the number of capture cycles in the multi-cycle scan-based test. The scan flip-flops without feedback, $R_{wof}$, are stitched by designer as a separate scan-chain for security considerations. {\em IBM BladeCenter$\textsuperscript{\textregistered}$ Cluster} with abort limit of 1 week is used. {\em '\bluecheck' $ $ is secure and '\redx' $ $ is insecure.}
\end{tablenotes}
\end{threeparttable}
\vspace{-1.5em}
\end{table*}

\vspace{-.5em}
\subsection{Resilience of scan-unrolled versions of {\em SeqL}-locked design to state-of-the-art attacks on logic locking}
\vspace{-.25em}
Table~\ref{tab:results-sota-attacks} confirms the resilience of {\em SeqL}-locked design to state-of-the-art attacks on logic locking like {\em Double-DIP} (DDIP)~\cite{double-dip}, {\em ScanSAT} (SS)~\cite{lilas:aspdac19}, {\em HackTest} (HT)-attack~\cite{hacktest},  functional-analysis-attacks on logic-locking (FALL)~\cite{pramod-faa}, and SMT-attack~\cite{smt-attack}. 
All experiments were run on IBM BladeCenter$\textsuperscript{\textregistered}$  Cluster, with an abort-limit of {\em 1 week}. Those entries in the table which are empty, correspond to all those cases which have crossed this abort-limit while performing key decryption. Similarly, for some cases the solver returns \texttt{No-key} (indicated as NK in the table). The resilience verification flow for oracle-guided attacks is similar to the flow in Figure~\ref{fig:flowchart}. 
For the oracle-less attacks, the resilience verification flow is slightly different because of absence of the oracle, however {\em lcmp} verifier is still used for formal-equivalence-checking.


\subsection{Resilience of SeqL vs. EFF against SAT-Attacks on sequential benchmarks}

Table~\ref{tab:results-proposed-eff-sequential} shows the results of applying the procedure shown in Figure~\ref{fig:flowchart} on ITC'99 open-source sequential gate-level benchmarks and flattened \texttt{RISC-V CPU} netlist. The \texttt{RISC-V CPU} RTL is obtained from~\cite{riscv-rtl}, and gate-level synthesis is performed using {\em Nangate\ 45nm} library using {\em Synopsys Design Compiler$\textsuperscript{\textregistered}$}. Scan chains and EDT-compression are inserted into the gate-level netlist using {\em Mentor Graphics TestKompress}$\textsuperscript{\textregistered}$ (decompressor and compactor will not be used because the attack is launched in EDT-bypass mode). 

The scan-inserted gate-level-verilog is converted to the bench format used in the attack tools, using an in-house {\em Python} script. 
The attack tools only support basic gates like {\em AND/OR/NAND/NOR/XOR/XNOR/NOT/BUF/MUX}, however the {\em RISC-V} gate-level-verilog contains more complex gates like {\em AOI} (and-or-invert), {\em OAI} (or-and-invert), {\em HA} (half-adder) and {\em FA} (full-adder). Our {\em Python} script internally converts each of these complex gates into a composition of basic gates, before final conversion to {\em bench}, which is acceptable because {\em IBLA} algorithm inserts scan-locks and does not affect combinational logic.  
The compression hardware is not converted because the attacks are meant to be launched in EDT-bypass mode. 
The columns \#{\em SFFs}, \#{\em SCs}, {\em Res.} and {\em Ov.} indicate number of scan flip-flops, number of scan-chains, resilience and overhead respectively. The resilience rate of {\em EFF} was 0\%, while that of {\em SeqL} was 100\%, thus indicating the superiority of  {\em SeqL} over {\em EFF}. An abort limit of {\em 1 week} was used for key decryption. 

\vspace{-.5em}
\subsection{Resilience to Multi-cycle attacks~\cite{guin:2018}}
\vspace{-.25em}

\label{sec:multi-cycle-tests}
So far, we discussed attacks using a single capture cycle. The attacker can also run the circuit for $N>1$ capture cycles (multi-cycle test), without affecting the shift cycles. 
This attack can be modeled by time-unrolling the reverse-engineered netlist as well as the oracle $N$ times.
Since scan-in and scan-out phases span hundreds of clock cycles and $N$ is in general relatively very small, running at slow-speed or at-speed will not significantly affect test-time/attack-time. 
Table~\ref{tab:results-proposed-eff-sequential} shows results for this attack. Similar to single-cycle attack ($N=1$), {\em SeqL} was resilient to multi-cycle attack ($N=2$, $5$) across all benchmarks. For $EFF$, since key is successfully recovered for $N=1$ itself, resilience results for $N>1$ were not shown.

\vspace{-.5em}
\subsection{Resilience to Shift-and-Leak attack (SaLa)~\cite{shift-and-leak-iccad19}}
\vspace{-.25em}
\label{sec:sala}
 {\em RDFS}~\cite{guin:2018} inserts special secure cells (SCs) into scan-chains to drive the key-gates. Unlike {\em RDFS}, {\em SeqL} key-gates are directly driven by the tamper-proof memory, without SCs in between. The first goal of {\em SaLa} is to find leaky cells, and shift the content of SCs into leaky cells. Due to absence of SCs in {\em SeqL}, this first goal is never achieved. The second goal of {\em SaLa} is to find the {\em leak condition} and satisfy it. Since {\em SeqL} locks the scan-chain itself, it is mandatory to know the scan-key upfront to invoke test generator and find the {\em leak condition}. Since the goal is itself key-decryption, it is not possible to find the {\em leak condition}, let alone satisfy it. Thus, {\em SeqL} is inherently resilient to {\em SaLa}.

\begin{table}[!t]
\renewcommand{\arraystretch}{0.78}
\setlength\extrarowheight{2pt}
\begin{center}
\caption{Area, Timing and Energy Overhead Comparison}
\vspace{-0.1in}
\label{tab:area-timing-energy}
\begin{tabular}{|c|c|c|c|c|c|c|}
\hline
$FF$ & \# Ts  & $T_{setup}$ & $T_{CK-to-Q}$ & \% Inc. &  EPT	 & \% Inc. \\
\hline
Orig. & $38$ & $45ps$ & $113ps$ & - &  $13.1 fJ$ & - \\
\hline 
$EFF$    & $48$ & $45ps$ & $163ps$ & $44\%$ & $17.1 fJ$ & $31\%$\\
\hline 
$SeqL$   & $50$ & $45ps$ & $127ps$ & $12\%$ & $13.9 fJ$ & $6\%$\\
\hline
\end{tabular}
\end{center}
\vspace{-1.5em}
\end{table}

\subsection{Overheads}
\vspace{-.25em}
\label{sec:overheads}
 Table~\ref{tab:area-timing-energy} shows the comparison of area, timing and energy for original, $EFF$-style and proposed $SeqL$-style locked scan flip-flops, obtained using SPICE transistor-level simulation. $NGSPICE$ open-source simulator, $Nangate\ 45nm$ library scan flip-flop and $45nm$ predictive technology model was used to arrive at these results. From Table~\ref{tab:area-timing-energy}, it is evident that the proposed $SeqL$ flip-flop has $22\%$ and $19\%$ reduction in $T_{CK-to-Q}$ and Energy-Per-Toggle ($EPT$) respectively, with only $4\%$ area overhead as compared to $EFF$ flip-flop.

\section{Conclusions} 
\label{conclusion}
\vspace{-0.25em}
We have proposed {\em SeqL}, that performs functional isolation and FI-SQ locking. {\em SeqL} hides a major fraction of the functionally correct keys, thus maximizing functional output corruption. We have shown both the theoretical and empirical improvements in the security of scan-locking. The results have shown 100\% resilience to state-of-the-art oracle-guided as well as oracle-less attacks. Furthermore, since combinational key (excluding FIs) is completely recovered, it is sufficient to lock FI-SQ pairs, making {\em SeqL} cost-efficient. Moreover, we have demonstrated {\em SeqL} on large designs such as \texttt{RISC-V CPU}, demonstrating its applicability in mainstream industry practice.

\section{Acknowledgements}
This work is supported in part by the German Research Foundation (DFG), Cluster of Excellence ”Center for Advancing Electronics Dresden”, Technische Universitat Dresden.


\vspace{-.5em}
\bibliographystyle{IEEEtran}
\bibliography{refs}

\end{document}